\documentclass[apjl]{emulateapj}

\shorttitle{Probing Curvature Effects in The Fermi GRB 110920}
\shortauthors{Shenoy et al.}

\begin{document}

\newcommand{\e}{\epsilon}
\newcommand{\ep}{\epsilon^\prime}
\newcommand{\et}{\epsilon_T}
\newcommand{\epk}{\epsilon_{pk}}
\newcommand{\tp}{t^\prime}


\title{Probing Curvature Effects in The Fermi GRB 110920}


\author{A. Shenoy\altaffilmark{1}, E. Sonbas\altaffilmark{2,3}, C. Dermer,\altaffilmark{4}, L. C. Maximon\altaffilmark{1}, K. S. Dhuga\altaffilmark{1},
P. N. Bhat\altaffilmark{5}, J. Hakkila\altaffilmark{6}, W.~C.~Parke\altaffilmark{1}, G. A. Maclachlan\altaffilmark{1}, T. N. Ukwatta\altaffilmark{7}}
\affil{$^1$Department of Physics, The George Washington University, Washington, DC 20052, USA}
\affil{$^2$University of Adiyaman, Department of Physics, 02040 Adiyaman, Turkey}
\affil{$^3$NASA Goddard Space Flight Center, Greenbelt, MD 20771, USA}  
\affil{$^4$Space Science Division, Code 7653, Naval Research Laboratory, Washington, D.C. 20375, USA}
\affil{$^5$CSPAR, University of Alabama in Huntsville, Huntsville, AL 35805, USA}
\affil{$^6$Department of Physics and Astronomy, College of Charleston, Charleston, S.C. 29424, USA}
\affil{$^7$Department of Physics and Astronomy, Michigan State University, East Lansing, MI 48824, USA}
\email{ashwinsp469@gmail.com}



\begin{abstract}
Curvature effects in Gamma-ray bursts (GRBs) have long been a source of considerable interest. In a collimated 
relativistic GRB jet, photons that are off-axis relative to the observer arrive at later times than on-axis photons and 
are also expected to be spectrally softer. In this work, we invoke a relatively simple kinematic two-shell collision 
model for a uniform jet profile and compare its predictions to GRB prompt-emission data for observations that have been 
attributed to curvature effects such as the peak-flux--peak-frequency relation, i.e., the relation between the 
$\nu$F$_\nu$ flux and the spectral peak, E$_{pk}$ in the decay phase of a GRB pulse, and spectral lags. In addition, we 
explore the behavior of pulse widths with energy. We present the case of the single-pulse Fermi GRB 110920, as a test for 
the predictions of the model against observations.
\end{abstract}



\section{Introduction}

Pulses in GRB light curves are thought to be produced by collisions between relativistic shells ejected from an active central 
engine  (Rees and Meszaros 1994). The interception of a more slowly moving shell by a second shell that is ejected at a later 
time, but with greater speed, produces a shock that dissipates internal energy and accelerates the 
particles that emit the GRB radiation. This scenario is widely adopted in order to model pulses in GRB light curves 
(e.g., Daigne and Mochkovitch\ 1998; Zhang et al.\ 2009). Studies of pulses are important to determine if GRB sources require 
engines that are long-lasting or impulsive, and to determine the likely radiation mechanism(s), with important implications 
for the nature of the central engine.\\

\indent Spectral lags, where low energy photons reach the observer at later times than high-energy photons, are seen in 
a significant fraction of GRBs. Cheng et al.~(1995) were the first to analyze the spectral lag of GRBs, which 
they determined as the time delay between the peaks in the Burst and Transient Source Experiment (BATSE) Large Area Detector (LAD) 
channel 1 (25 - 50 keV) and channel 3 (100 - 300 keV) light curves. Since then, several authors have analyzed spectral lags in GRBs, while 
also extending these observations to the Swift and Fermi GRB samples (e.g., Norris et al.\ 1996; Norris, Marani \& Bonnell 2000, Wu \& Fenimore 2000; Chen et al.\ 2005; Ukwatta et al.\ 2010; Sonbas et. al.\ 2012) The leading model to explain the spectral lag is the curvature effect, i.e.\ the kinematic effect 
due to the observer looking at an increasingly off-axis annulus area relative to the line-of-sight (Fenimore et al. 1996; Salmonson 2000; Kumar \& Panaitescu 2000; Ioka and Nakumura 2001; Qin 2002; Qin et al. 2004; Dermer 2004; Shen et al. 2005; Lu et al. 2006). Softer low-energy radiation comes from the 
off-axis annulus area due to smaller Doppler factors. This radiation is also delayed at the observer end with respect to 
on-axis observation due to the geometric curvature of the shell.\\ 

\indent The existing models as well as observations suggest that a connection exists between the observed hard-to-soft 
spectral evolution of GRB pulses and spectral lags. It is therefore important to understand the mechanism that produces this 
evolution. Tavani (1996) proposed that this hard-to-soft spectral evolution is caused by the variation of the average Lorentz factor 
of pre-accelerated particles and the strength of the local magnetic field at the GRB site as the synchrotron emission evolves 
within the burst. Liang (1997) proposed a physical model of hard-to-soft spectral evolution in which impulsively accelerated 
non-thermal leptons cool by saturated Compton up-scattering of soft photons. Kocevski \& Liang (2003) have analyzed a sample of 
19 GRBs and  found a positive correlation between the decay rate of the peak energy and the spectral lag. Ukwatta et al.\ (2010) have 
analyzed a sample of 31 Swift GRBs with known red-shifts and determined the spectral lags between fixed source frame bands, 100 -- 150 keV and 200 -- 250 keV. They also determined that the source-frame E$_{pk}$ lies beyond the higher-energy band, 100 -- 250 keV for a majority of these bursts. Based on this result, they suggest that spectral evolution may not be the dominant process causing the observed spectral lag.\\

\indent Borgonovo and Ryde (2001) studied the spectral evolution in the prompt-emission in GRBs  by performing a time-resolved 
spectral analysis of BATSE single pulses and showed that in many cases the $\nu F \nu$ flux at the E$_{pk}$ for each time segment 
was $\propto$ E$_{pk}^\eta$ (hereafter referred to as the peak-flux--peak-frequency relation), with $\eta$ ranging from $\approx 0.6$ to $3$. 
The exponent $\eta$ was found to stay roughly constant for pulses within the same GRB. Dermer (2004) modeled and analyzed GRB pulses based 
on curvature effects with a Broken-Power-Law (BPL) rest-frame spectrum and showed that in the curvature limit, $\eta$ was equal to 3 for 
pulses with a wide range of temporal properties. \\

\indent While several studies (Qin 2002; Qin et al. 2004; Shen et al.\ 2005; Lu et al.\ 2006) have been performed to determine the 
role played by curvature effects, both in the spectral evolution of GRB prompt-emission as well as  in producing spectral lags, a number of questions remain unanswered. Specifically, we note that the dependence of the lag on the radius at which the emission takes place is quite unclear at this stage with seemingly contradictory results being reported in the literature (Shen et al.\ 2005; Lu et al.\ 2006). While that is not the focus of our current study, it is a matter of considerable interest and will be the central topic of a forthcoming work. In this work we focus on the role played by the thickness of colliding shells on observables such spectral lags, the  $\nu F \nu$ flux vs.  E$_{pk}$ relation, and the behavior of the corresponding pulse widths as a function of energy. Other studies have also attempted to determine the role played by the evolution of the rest-frame spectrum on the observables such as the spectral lags and the evolution of the pulse-widths with energy within the context of a curvature model(Qin et al.2005; Lu et al. 2006; Qin et al. 2009; Peng et al. 2011). Again, such considerations are very important but we do not specifically consider the effects of the evolution of the rest-frame spectrum in this paper.\\

\indent The paper is organized as follows: the basic features of the model are presented in section 2, followed by a description of the sample selection criteria, analysis methodology, and a case study in section 3. The discussion of our main results is presented in section 4, followed by a summary 
of our conclusions in section 5.
 \begin{table*}
 \centering
\caption{Selected parameters for the generation of the model light curves unless otherwise stated.}
  \begin{tabular}{lcccccccccc}
\hline
$\eta_r$	& $\eta_t$ & $\eta_\Delta$ &  $t_{var} (s)$ & $\Gamma$ & z & $\theta_{jet}$ & E$_{pk,0}$ (keV) & $d_L$ (cm)  & $u_0$ (ergs cm$^{-3}$)\\
\hline
1.0 & 1.0 & 1.0 & 1.0 & 300 & 1.0 & 4.0/$\Gamma$ &250.0 & $2.2\times 10^{28}$  & 1.0\\
\hline
\end{tabular}
\end{table*}

\section{The Model} 
We have used a particular representation of the internal shock model for our purposes (Dermer 2004). This model consists of a 
single two-shell collision event occurring at a radius $r_0$ from the source, generating a uniform spherical shell, with Lorentz factor 
$\Gamma$, that radiates for a co-moving time between $t^{\prime}_0$ and $\tp_0 + \Delta\tp$ with $\Delta\tp =  \eta_t \Gamma t_{var}/(1+z)$, where $t_{var}$ is the observed variability time scale. The rest-frame--emission profile is assumed to be rectangular with instantaneous rise and decay phases. The opening angle of the jet is assumed to be $4/\Gamma$. Emission from angles greater than $4/\Gamma$ are ignored. This is a suitable compromise between considering emissions from an entire fireball surface ($0 < \theta < \pi/2$) and a collimated jet ($\theta \sim 1/\Gamma$). As noted by Qin et al. 2004, limiting the radiation to $\theta < 1/\Gamma$ leads to a cutoff-tail problem whereas the contributions from areas at $\theta > 1/\Gamma$ fall off very rapidly. We find such a choice to be suitable for producing pulse profiles that may be directly compared with observations. The co-moving width of the shell $\Delta r^\prime$ is assumed to remain constant during the period of illumination and given by $\Delta r^\prime = \eta_{\Delta} \Gamma~c ~t_{var}/(1+z)$. It is in this respect that the chosen model differs from previous studies. Previous studies (see for instance Qin et al. 2004, Shen et al. 2005) have studied the effects of curvature from a spherical surface in great detail. As shown subsequently, the effect of a finite shell where $\Delta r^\prime$ is comparable to $c \Delta\tp$, has a significant effect on the predicted observables such as spectral lags and pulse widths as a function of energy when compared to the models that study curvature effects from a surface. As $\Delta r^\prime << c \Delta\tp$, we approach the infinitesimal shell  or emission-surface situation and are able to recover many of the predictions of the aforementioned models. The emission spectrum in the co-moving frame may be described by any suitable spectral function such as a Broken Power-Law (BPL), Band, or Comptonized--E$_{pk}$, peaking at a co-moving photon energy E$^\prime_{pk,0}$  =  (1 + z) E$_{pk,0}$/2$\Gamma$ where E$_{pk,0}$ is the observer frame E$_{pk}$ at the start of the pulse. The spectral indices at E  $< $ E$_{pk,0}$  and at E $ >$ E$_{pk,0}$  in counts space are denoted $\alpha$ and $\beta$, respectively. The curvature constraint requires that $r \lesssim 2\Gamma^2 c~t_{var}/(1+z)$ (Fenimore et al. 1996). The radius is thus written using the expression $r_0  = 2\eta_r \Gamma^2 c~t_{var}/(1+z)$, with $0 \lesssim \eta_r  \lesssim 1$. The parameters $\eta_t$, $\eta_{\Delta}$ and $\eta_r$ thus control the blast-wave duration, shell-thickness and radius of emission respectively.\\

\begin{figure}
\epsscale{1.20}
\centering
\plotone{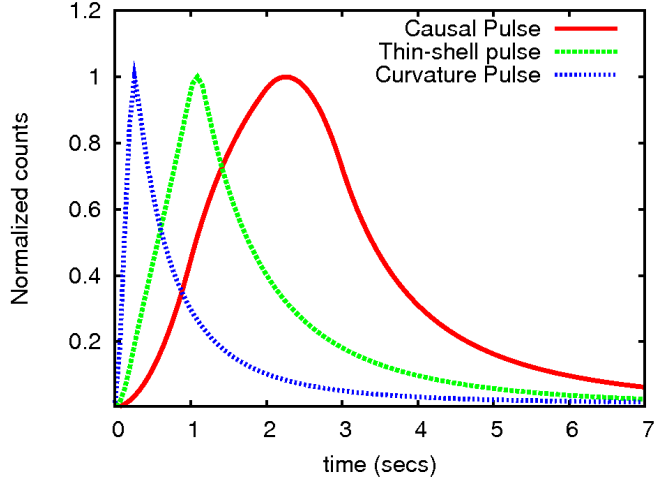}
\caption{Normalized light curves obtained using selected parameters except: Thin-shell pulse, $\eta_\Delta = 0.1$; and the Curvature pulse,  $\eta_\Delta =  \eta_t = 0.1$. The case of $\eta_t = \eta_\Delta = \eta_r = 1$ is referred to as a Causal pulse (see Dermer 2004 for a detailed discussion of these three generic types of pulses).}
\label{fig1}
\end{figure}

\begin{figure}
\epsscale{1.20}
\centering
\plotone{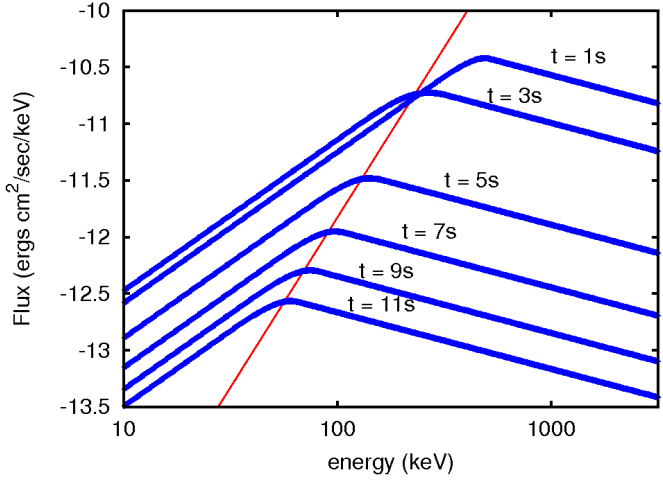}
\caption{Evolution of the spectral energy distribution due to curvature effects for the case of the causal pulse in Fig.\ 1. 
In the declining phase of the pulse, the value of $f_{E_{pk}} \propto$ E$_{pk}^3$ is shown by the red line.}
\label{fig2}
\end{figure}

\begin{figure}
\epsscale{1.20}
\centering
\plotone{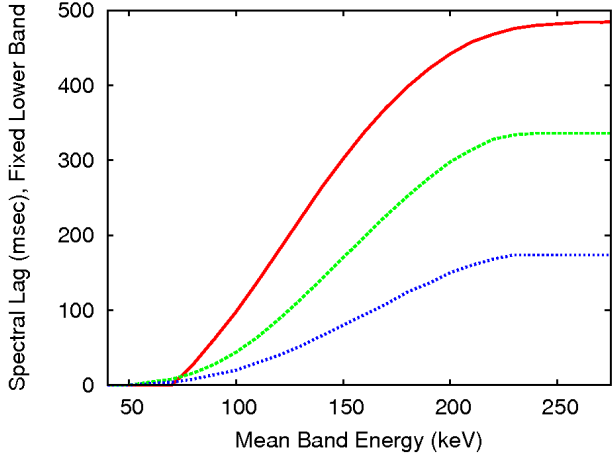}
\caption{Lag vs. Energy for selected parameters except shell thickness $\eta_\Delta$. Red: $\eta_\Delta = 1.0$;  
Green: $\eta_\Delta = 0.5$;  Blue: $\eta_\Delta = 0.1$.}
\label{fig3}
\end{figure}

\begin{figure}
\epsscale{1.20}
\centering
\plotone{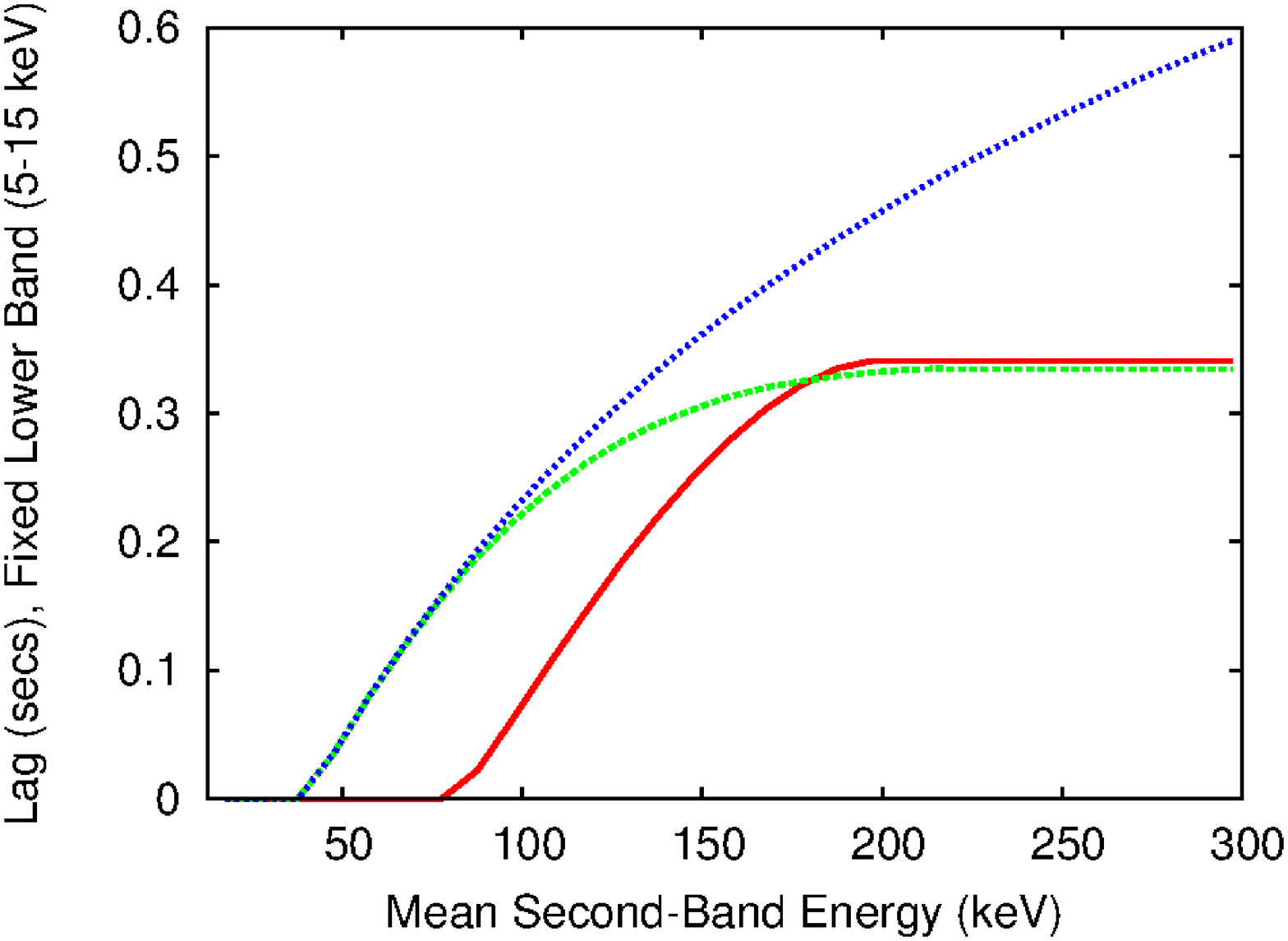}
\caption{Lag vs. Energy due to different rest-frame spectral functions with selected parameters except E$_{pk,0}$ = 200 keV. Here, Red: Broken-power-law with $\alpha = -1/3$, $\beta = -2.5$, Green: Band function with $\alpha = -0.8$, $\beta = -2.25$ and  Blue: Comptonized-E$_{pk}$ with $\alpha = -0.8$.}
\label{fig4}
\end{figure}

\begin{figure}
\epsscale{1.20}
\centering
\plotone{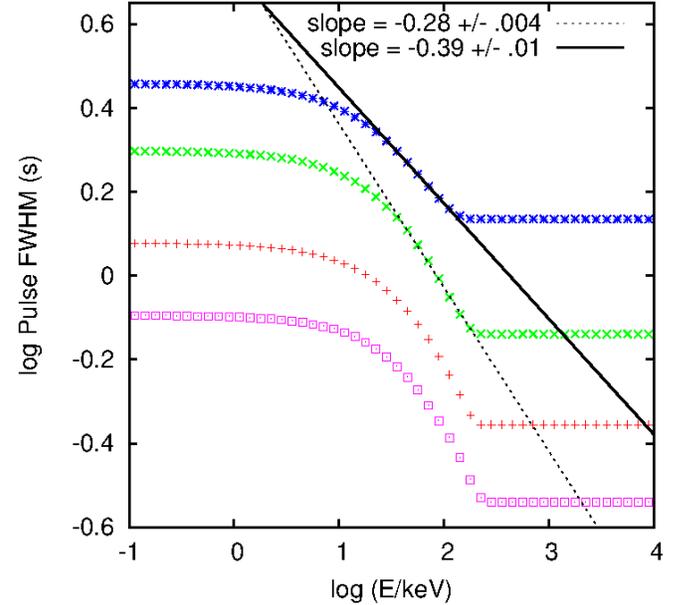}
\caption{Pulse FWHM vs. Energy for different shell-collision parameters $\eta_\Delta$ and $\eta_t$ for selected parameters except E$_{pk} = 200$ keV, a Band spectrum with $\alpha = -1.0$ and $\beta = -2.25$ and $t_{var} = 1.0$ s, with: Pink: A pulse with Infinitesimal duration and shell thickness with $\eta_\Delta = 0.001$ and $\eta_t = 0.001$ Red: Curvature pulse with $\eta_\Delta = 0.1$ and $\eta_t = 0.1$; Green: Thin-shell pulse with $\eta_\Delta = 0.1$ and $\eta_t = 1$; and Blue: Causal pulse with $\eta_\Delta = 1$ and $\eta_t = 1$. Also shown is the exponent of the power-law that best fits the model points for the thin-shell and Causal pulses.}
\label{fig11}
\end{figure}   

Unless otherwise stated, the selected parameters used for the numerical calculations of the light curves, spectra and spectral lags are shown in Table.\ 1. Fig.\ 1 shows the normalized, generic pulse shapes obtained using the selected parameters. As can be seen, the cases of the thin-shell pulse ($\eta_\Delta << \eta_t$) and the curvature pulse ($\eta_\Delta = \eta_t << \eta_r$) produce the sharp featured light curves noted by Qin et. al. 2004 for emission from a  rectangular pulse profile in the co-moving frame, and which are attributed to the effect of a suddenly-dimming emission profile. These two cases most closely correspond to a long duration and a short duration pulse in the co-moving frame respectively, and where the effects of a finite shell are suppressed. The case of the causal pulse ($\eta_\Delta = \eta_t = \eta_r = 1$) however, shows that emission from a finite shell can produce smooth light curves without the need for a slowly dimming co-moving emission profile. Fig.\ 2  shows the evolution of the spectral energy distribution for the case of the curvature pulse shown in Fig.\ 1. The $\nu$F$_\nu$  peak flux $f_{E_{pk}}$ $\propto$ E$_{pk}^3$ equality line is shown in the decay portion of the pulse. Dermer (2004) shows that this equality holds in the declining phase for all pulses (a similar result has been derived by Qin et al. 2009). We also note that the presence of a finite shell affects the low energy and high energy fluxes equally and therefore does not effect the shape of the spectrum as a function of time as first noted by Qin et. al. 2002 in the context of emission from a fireball surface. Fig.\ 3 shows the spectral lag as a function of energy for 
a case with selected parameters but with varying shell thickness parameter, $\eta_\Delta$. Fig.\ 4 shows the lag as a function of energy for a case with selected parameters but with different rest-frame spectral functions (BPL, Band, and Comptonized--E$_{pk}$). Fig.\ 5 shows the pulse Full-width at half-maximum (FWHM) as a function of energy for the various profiles shown in Fig. 1. For the purpose of comparison we have used a Band function with $\alpha = -1.0$ and $\beta = -2.25$, identical to Qin et. al. 2005. These authors have shown that the Doppler effect of a relativistically expanding fireball could lead to a power-law trend for the pulse width as a function of energy within a certain energy range. By taking a sizable sample of BATSE GRBs, they demonstrated that the pulse widths exhibit a plateau/power-law/plateau feature as a function of energy. They also note that the power-law index depends strongly on E$_{pk}$ and the rest-frame radiation spectrum. The plateau/power-law/plateau feature reported by Qin et al. 2005 is well reproduced here. Furthermore, we note that the power-law exponent is sensitive to the assumed thickness of the shell.

\begin{figure}
\epsscale{1.20}
\centering
\plotone{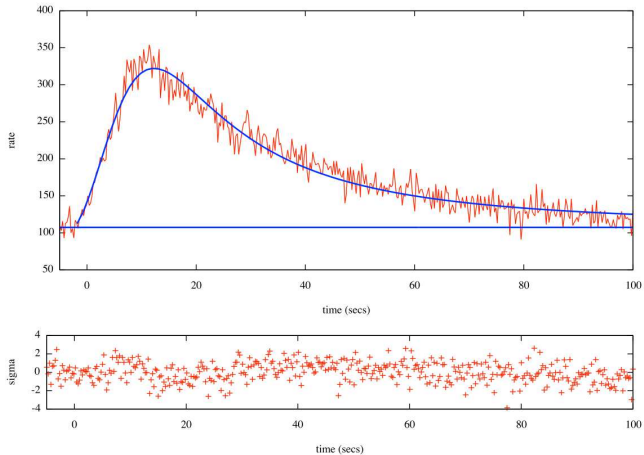}
\caption{KRL pulse-fit for the light curve for GRB 110920 with residuals.}
\label{fig5}
\end{figure}

\begin{figure}
\epsscale{1.20}
\centering
\plotone{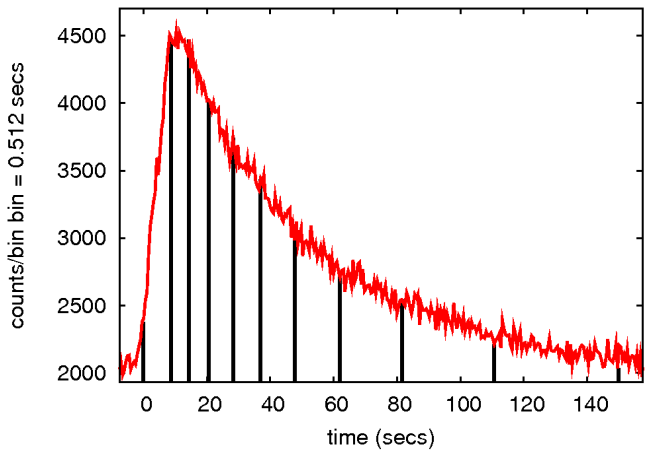}
\caption{Light curve segments for 110920 with equal fluences.}
\label{fig6}
\end{figure}

\begin{figure}
\epsscale{1.20}
\centering
\plotone{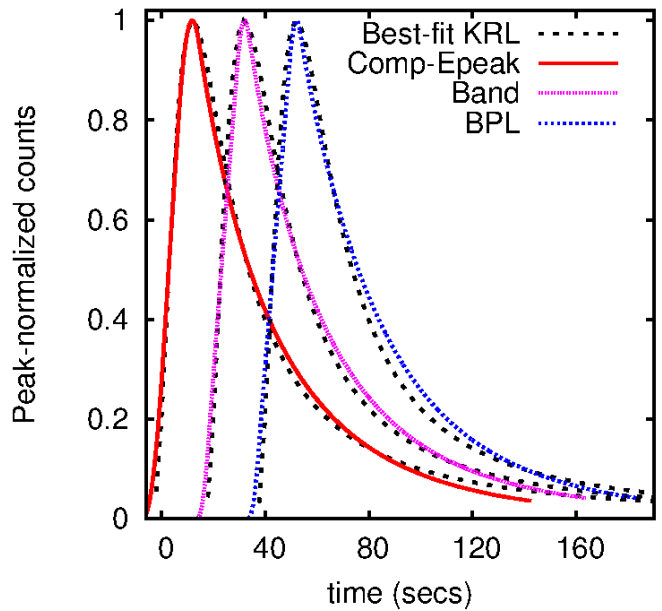}
\caption{Comparable pulses generated using the BPL (E$_{peak} = 300$ keV, $\alpha = -\frac{1}{3}$, $\beta = -2.5$),
 Band (E$_{peak} = 334$ keV, $\alpha = -0.2$, $\beta = -2.65$)and Comptonized--E$_{pk}$ (E$_{peak} = 280.1$ keV, $\alpha = -0.46$) spectral functions. 
 The light curves have been offset by 20 seconds for better viewing.}
\label{fig7}
\end{figure}

\begin{figure}
\epsscale{1.20}
\centering
\plotone{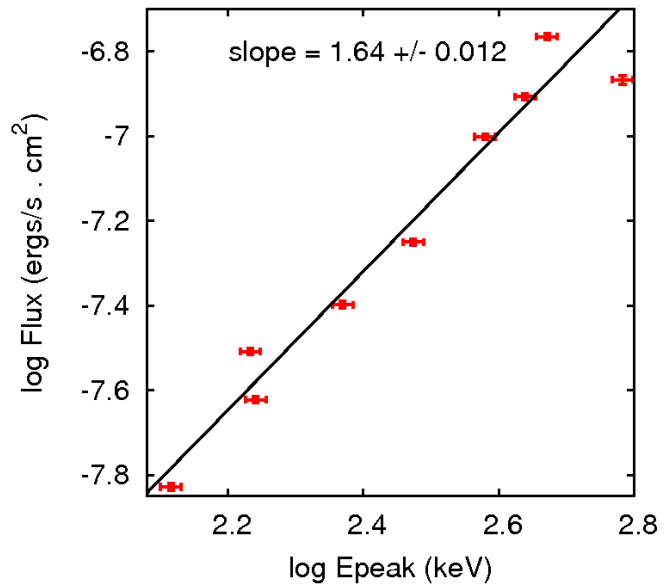}
\caption{$\nu F \nu$ Flux vs. E$_{pk}$ for the data from GRB 110920. The data were fit with the best-fit Comptonized--E$_{pk}$ function in the range 100-985 keV.}
\label{fig8}
\end{figure}

\begin{figure}
\epsscale{1.20}
\centering
\plotone{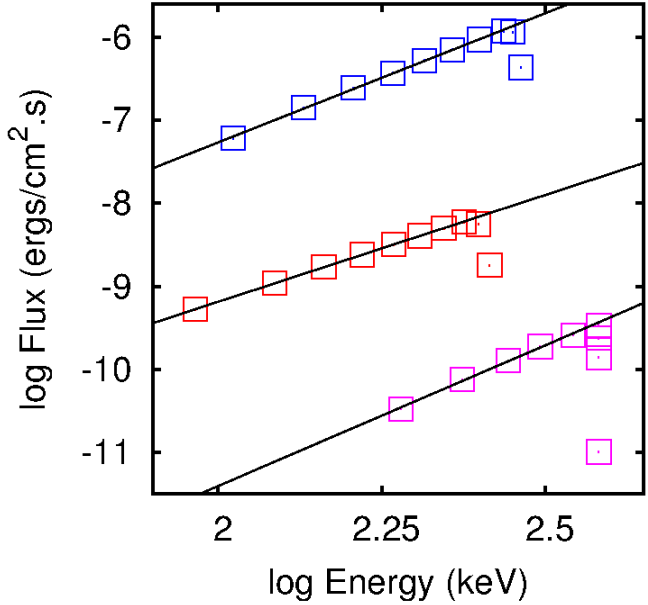}
\caption{ $\nu F \nu$ Flux vs. E$_{pk}$ from the model for the three rest-frame spectral functions described in the text with Blue: BPL; slope = 3.11 +/- 0.04, 
Pink: Band; slope = 3.39 +/- 0.10  and Black: Comptonized--E$_{pk}$; slope = 2.57 +/- 0.01, for time segments identical to those used for the data. The flux scale has been offset for better viewing.}
\label{fig9}
\end{figure}

\begin{figure}
\epsscale{1.20}
\centering
\plotone{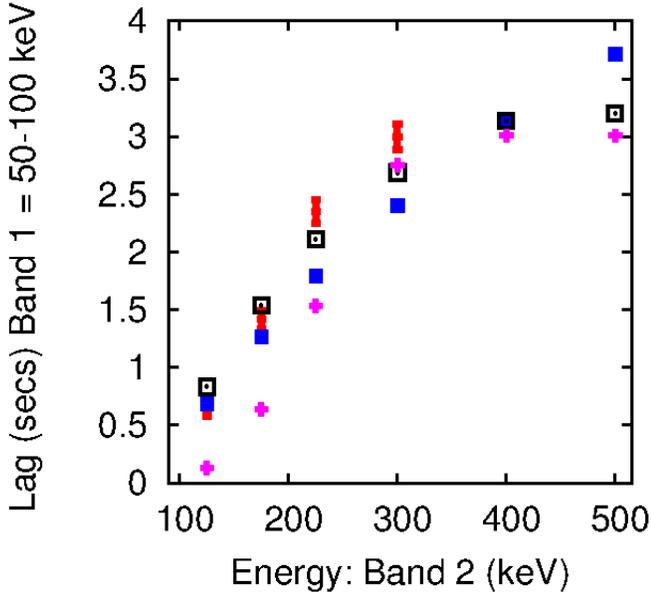}
\caption{Lag vs. Energy from the model for the pulses shown in Fig.~7 with Red: Data, Solid blue squares: Comptonized--E$_{pk}$, Hollow black squares: 
Band function and Pink crosses: BPL. The Band 2 energies are the mid-point of the energy in the second band.}
\label{fig10}
\end{figure}

\begin{figure}
\epsscale{1.20}
\centering
\plotone{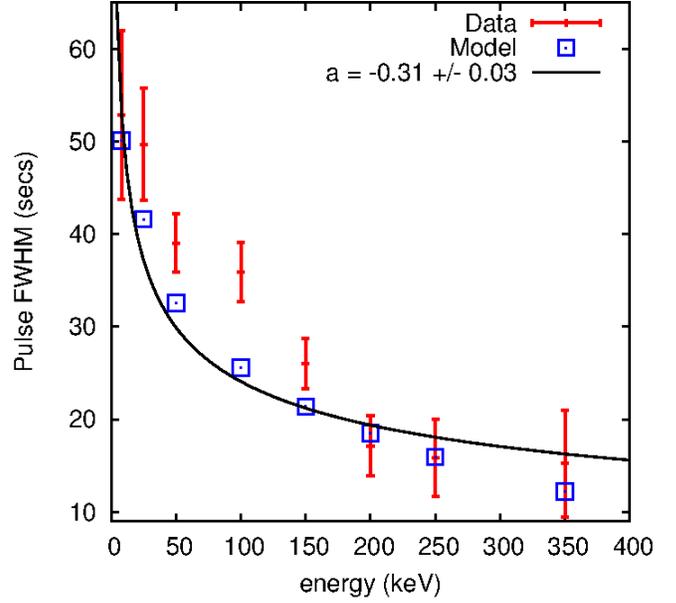}
\caption{Pulse FWHM vs. Energy for the data and the model for GRB 110920 for identical energy bands. Also shown is the exponent of the power-law that best fits the model points.}
\label{fig11}
\end{figure}

\section{Sample Selection and Methodology.}
As a first step we analyze either single-pulse GRBs, or GRBs with relatively simple light curves where the individual pulses within 
a multi-pulse structure in the light curve can be distinguished. In addition, we require that the GRB pulses be bright enough and of sufficient 
duration (the duration of the pulse is particularly relevant to tests of the peak-flux -- peak-frequency relation) so that we may obtain 
reliable results from our analyses.\\

\indent After identifying a potential candidate GRB,  we pulse-fit the GRB light curves using a suitable pulse function (such as the Kocevski-Ryde-Liang (KRL) pulse function: see Kocevski  Ryde \& Liang 2003 or the Norris pulse function: see Norris et al. 2005) in multiple energy bands. In order to support the supposition that a given strong pulse (obtained from a suitable pulse-fit) is not made from overlapping multiple pulses (within statistics), we perform a wavelet based minimum-variability time-scale (MTS) extraction. In essence, the MTS is a measure of the smallest temporal structure in a lightcurve. The full details of its extraction, and the technique in general, are given in MacLachlan et al. (2013). The correlation between MTS and pulse properties such as rise times and widths is discussed in MacLachlan et al. (2012). The best-fit pulse profile is then used as a representation of the light curve from the data.\\

\indent The time-integrated spectrum of the GRB is fit with a suitable function (Band, Comptonized--E$_{pk}$  etc.). The best-fit spectral function is used as the rest-frame emission spectrum in the model. The model parameters are then varied to generate a light curve that best matches the best-fit pulse profile. The light curve is subdivided into time segments with equal, and sufficiently high background-subtracted fluence in order to minimize the effects of varying signal-to-noise and an E$_{pk}$  is extracted via a spectral fit for each time segment. The $\nu$F$_\nu$ flux is extracted at E$_{pk}$ in a range spanned by the E$_{pk}$-error. The model light curve is treated in an identical fashion as the data with regard to segments. Model fluxes and E$_{pk}$'s are extracted 
and the peak-flux -- peak-frequency relation is tested. In addition, we extract spectral lags in suitable, identical energy bands from the data and the model and compare the predicted and the observed spectral-lag-energy evolution. The spectral lags are extracted using the cross-correlation-function analysis method as described in Ukwatta et. al (2010).\\

\subsection{GRB 110920 - A Test Case}
The Fermi GRB 110920 is a single-pulse burst with a relatively long fast-rise, exponential-decay structure with a $T_{90}$ of 170 $\pm$ 17 seconds.  The best fit Band parameters (see McGlynn et al. 2012 for a detailed discussion on the properties of this GRB) for the time interval [$T_0+0.003,T_0+52.737$]  (where $T_0$ is the trigger time) were $\alpha = -0.20 \pm 0.02$, $\beta = -2.65^{+0.07}_{-0.09}$ and $E_{peak} = 334 \pm 5$ keV with C-stat = 3206.5 (485 d.o.f.). However, when a blackbody component was included in the fit, the C-stat was reduced to 2848.3 (483 d.o.f.). The peak energy of the Band component was shifted up to 
$E_{peak} = 978^{+154}_{-121}$ keV and the temperature of the blackbody was found to be $kT = 61.3^{+0.7}_{-0.6}$ keV. The low energy index $\alpha$ became ($-1.05 \pm 0.04$).  McGlynn et al.~(2012) have attributed this blackbody component to the photospheric emission (see for instance: Ryde, 2005; Rees \& Meszaros, 2005). Since the redshift (z) is undetermined for this GRB, they have assumed a value of z = 2 and then used photospheric emission models to obtain a bulk Lorentz factor, $\Gamma$ of $\sim$ 440.\\

\indent A careful analysis, based partly on the results presented above, allowed us to separate the underlying pulse structure from the overall structure of the light curve. Fig.~6 shows the full light curve for the GRB in the energy range 100 -- 985 keV along with the best-fit KRL pulse function. The energy range was so chosen to avoid contamination from the soft component (below 100 keV).\\ 

\indent The best-fit KRL function was then used as the representation of the light curve to be matched by the model. We assumed a Bulk Lorentz factor $\Gamma$ = 440 and z = 2 (as in McGlynn et al.~2012). Fig.\ 7 shows the segmentation of the light curve into equal-fluence segments. The time-integrated energy-spectrum of the pulse was fit with a Band function (as in McGlynn et al~2012, in the energy range, 8 - 985 keV) and a Comptonized--E$_{pk}$  function (in the range 100-985 keV) and these, along with a theoretical BPL function (originally used by Dermer 2004) were used as rest-frame spectra for the model. We varied the shell-collision parameters and extracted best-fit pulses using these three spectral functions (see Fig.\ 8 for the pulses as well as details of the corresponding spectral parameters). The resulting best-fit shell-collision parameters, together with a best-fit value of $t_{var} = 44$ seconds and the chosen values of $\Gamma$ and z, yielded a radius, $r_0$ of 1.5 $\times 10^{17}$ cm, a shell thickness, $\Delta r^\prime$ of 8.8 $\times 10^{12}$ cm and a co-moving frame pulse-duration, $\Delta t^\prime$ of 2.0 $\times 10^{3}$ seconds  for the Comptonized--E$_{pk}$ spectral function. The corresponding values for the Band and BPL functions were very similar. Figs.\ 9 and 10 show our results for the peak-flux--peak-frequency relation for the data with the best-fit Comptonized-E$_{pk}$  function and the model using the three different rest-frame spectra. Fig.\ 11 shows the spectral lags for the light curves of Fig.\ 8.\\ 

In order to explore the evolution of the pulse width with energy, we extracted the FWHM and plotted this as a function of energy. The plot is shown in Fig.\ 12. The energy bands chosen (in keV) were  8--25, 25--50, 50--100, 100--150, 150--200, 200--250, 250--350, 350--985. These energy bands were so chosen as to ensure a sufficient number of counts in each energy band for a reliable measurement of pulse FWHM while also providing a sufficient number and spacing of bands to show the trend curve.  As the KRL function does not fit the pulses below ~150 keV accurately, we employed a Monte-Carlo simulation where 1000 light curves were simulated in each energy band using the square-root of the counts as their errors (assuming independent Poisson distributions for the counts), the pulse FWHM was extracted for each light curve, and the mean and standard deviation of the 1000 pulse FWHMs were used as the pulse FWHM and its error respectively for each band. We fit a power-law of the form C$E^a$ to the model points and extracted an exponent of -0.31 +/- 0.03. 

\section{Discussion}

Before we turn to the test-case GRB, we note that in the model calculation for a broken-power-law rest-frame spectrum, the lag shows a relatively well-defined trend with no lags at energies below $\sim$ 0.3E$_{pk,0}$ and a constant lag for all energies above E$_{pk,0}$ (Fig.\ 4). Increasing or decreasing the value of E$_{pk,0}$, while keeping all other model parameters fixed, only shifts the entire curve in the direction of increase or decrease. Shen et al. (2005) studied the lags due to curvature effects using different rest-frame emission profiles. They found that for an infinitesimal shell, a rectangular profile produces no lags. We find the situation to be quite different for the case of a finite shell thickness. In addition, as depicted in Fig 4, a change in the rest-frame spectrum also has a significant effect in the evolution of the spectral lag with energy. In the case of a Band or Comptonized--E$_{pk}$ function, the lags are small for energies small compared to E$_{pk,0}$, and the lags for a Comptonized--E$_{pk}$ function do not show a saturation energy. This is primarily because the Comptonized--E$_{pk}$ function varies monotonically at all energies and does not have a well-defined break energy. The value of $r_0$ obtained for the test case is consistent with estimates for an internal shock model (see for e.g. Hascoet et al. 2012). While the value of $\Delta r^\prime$ ($\sim 10^{13}$ cm) seems reasonable, it is difficult to infer the significance of its absolute magnitude in the context of the current analysis. The inclusion of a finite-shell-thickness component in our curvature model also produces relatively large lags ($\sim$ a few seconds) without a need for extreme physical parameters such as $\Gamma <$ 50 (as concluded by Shen et al. 2005), or a large local pulse width ($\sim 10^7$ seconds as concluded by Lu et al. 2006). We find that a rest-frame pulse duration, $\Delta t^\prime \sim 10^3$ seconds is sufficient to produce such lags.\\   

Our results for the test case show that an internal shock model with a rest-frame spectrum identical to that used to fit the data (the Comptonized--E$_{pk}$ and Band  functions), reproduces the observed pulse profile as well as the observed spectral lags. It was difficult to determine if there was a saturation energy present in the lag-energy-evolution (Fig.~11) as  there were insufficient counts to extract lags in higher energy bands. As noted above, a finite shell thickness can account for observed lags even for the case of a rectangular rest-frame emission profile.  A second Band function component with a peak shifted to ~1 MeV (when a blackbody component is included in the fit) would imply a 1 MeV break energy in the lag-energy  plot, and would also predict no lags (or small lags in the case of the Comptonized--E$_{pk}$) below $\sim$300 keV. This does not match the observations. It can be seen from Fig.~10 that the exponent in the peak-flux--peak-frequency relation is close to  3 in all cases for the model pulses shown in Fig.~8. This confirms the observations of Dermer (2004) that the exponent at times after the peak of the light curve is close to 3 even with different choices of rest-frame spectra. However, this does not match the exponent obtained from the data (Fig.~9).  While a connection may exist between the observed spectral evolution and the spectral lags, it appears that curvature effects alone cannot describe this connection and additional emission mechanisms may be needed. We note in passing that Guirec et al.\ (2012) have included a blackbody component to describe the temporal and spectral properties of a number of GRBs.\\

We have also explored the behavior of the pulse width with energy. As shown in Fig.\ 12, the pulse-width decreases with energy approximately as a power law. Both the data and the model predictions exhibit similar trends although the low-energy agreement is marginal. The exponent of the power law (-0.31 +/- 0.03) matches well the exponent extracted by Fenimore et. al. (1995), who analyzed a large sample of bright BATSE bursts and obtained an average power-law exponent of about -0.4.  A similar result was also obtained by Peng et al. 2006 who analyzed a sizeable sample of bright single pulses in BATSE GRBs. In addition, in a recent work based on an analysis of 51 long-duration FRED-like single-pulses from the BATSE data, Peng et. al. 2012 showed that the curvature effect combined with a Band rest-frame spectrum can explain the energy dependence of the pulse widths. Cohen et al.\ (1997) have suggested that such an exponent is consistent with a population of electrons losing energy via synchrotron radiation, a process for which the exponent is predicted to be -0.5.\\

\section{Conclusions}
We have used a simple two-shell collision model to investigate curvature effects in the prompt emission of GRBs. We have examined the effects of emission spectra such as the Band, the Comptonized--E$_{pk}$, and the BPL functions.  We have focused primarily on the peak-flux -- peak-frequency relation and the evolution of the spectral lags and the pulse widths with energy. We compare our model results with the results of similar models in the literature and also present a test case study of GRB 110920. \\
\\
We summarize our main findings as follows:
\begin{itemize}
\item We find that introduction of a finite shell thickness in the curvature formulation can produce smooth light curves, i.e., without a rapid transition from the rise to the decay portions even for the case of a rectangular rest-frame emission profile. As we approach the infinitesimal shell (surface) approximation, we recover the sharp featured light-curve profile of Qin et. al. 2004 for a rapidly dimming intrinsic emission profile.
\item While we agree with Shen et al. (2005) that an infinitesimal shell produces no discernible spectral lag using a rectangular emission-pulse-profile, 
we find the situation to be different for a shell of finite thickness i.e., a finite spectral lag can be produced even with a rectangular pulse profile;
\item The spectral lag evolution as a function of energy is quite sensitive to the type of rest-frame spectrum. For example, the Comptonized--E$_{pk}$ model does not appear to exhibit a saturation energy at which the spectral lags reach a plateau phase as in the case of the Band and the broken-power-law functions. We agree with Shen at al (2005) that the spectral lags seem to approach a maximum when E$_{pk}$ is near the high-energy channel used in extracting the lag. Most likely this simply reflects the break energy present in the assumed rest-frame spectrum (i.e., Band and BPL);  
\item All rest-frame spectral models tested exhibit the peak-flux -- peak-frequency relation although with exponents that differ from the 
predicted exponent of 3. The significance of this discrepancy is not clear at this stage and warrants further investigation;    

\item The peak-flux -- peak-frequency test for GRB 110920 yields an exponent of 1.64 +/- 0.012 compared to the theoretical one of 2.57 +/- 0.01 (with 
Comptonized--E$_{pk}$ as the rest-frame spectrum). We consider this discrepancy to be significant and the result to be in disagreement with the prediction 
based purely on effects of curvature. Similar conclusions were reached by Dermer (2004), Qin (2009) and Borgonovo and Ryde (2001);

\item Both the data (test GRB) and the model exhibit a very similar power-law trend for the pulse width with energy. The plateau/power-law/plateau feature noted by Qin et al. 2005 is well reproduced with a given choice of key model parameters. In addition, we note that the power-law exponent is sensitive to the assumed shell thickness. For the test-case GRB, the power-law exponent matches well with exponents extracted from a larger sample of GRBs from earlier studies (Fenimore et al. 1995; Peng et. al. 2006, 2012); and  

\item Relatively good agreement is obtained with all rest-frame spectral models for the spectral lag versus energy for the test GRB. This is somewhat surprising given the result of the peak-flux -- peak-frequency test noted above. It would seem that some complex interplay is at work between various model parameters such as shell thickness, variability time scale, the energy evolution of E$_{pk}$ and the Lorentz factor. The investigation of the dependencies of these various parameters is ongoing.
\end{itemize}

The role of the reported soft component of the light curve for GRB 110920 has not been fully investigated in this study and is worth pursuing, particularly with regard to the behavior of the peak-flux--peak-frequency relation. Finally, we note that these studies are being extended to a larger sample of GRBs.

\section*{Acknowledgements} The authors (AS and KSD) would like to acknowledge A. Eskandarian  and O. Kargaltsev (both from the George Washington University) for their valuable contributions to the discussions as well as the financial support provided by them to A. Shenoy at various stages of this work.

\end{document}